\documentclass{ifacconf}

\usepackage[utf8]{inputenc}
\usepackage[T1]{fontenc}
\usepackage{breqn}
\usepackage{amsmath}
\usepackage{amsfonts}
\usepackage{amssymb}

\newtheorem{assumption}{Assumption}

\newtheorem{problem}{Problem}
\newtheorem{remark}{Remark}

\usepackage{enumitem}
\usepackage{tikz}
\usetikzlibrary{shapes,arrows,calc,fit}
\tikzset{
        block/.style = {draw, rectangle,
            minimum height=1cm,
            minimum width=2cm},
        input/.style = {coordinate,node distance=1cm},
        output/.style = {coordinate,node distance=4cm},
        arrow/.style={draw, -latex,node distance=2cm},
        pinstyle/.style = {pin edge={latex-, black,node distance=2cm}},
        sum/.style = {draw, circle, node distance=1cm},
    }
\usepackage{pgfplots} 
\pgfplotsset{compat=newest} 
\pgfplotsset{plot coordinates/math parser=false} 
\newlength\figureheight 
\newlength\figurewidth 
\pgfplotsset{
    every axis plot post/.style={
        line join=round
    }
}

\usepackage{tikz}
\usepackage{tikzscale}

\definecolor{myorange}{cmyk}{0,0.35,0.85,0} 
\definecolor{mypurple}{cmyk}{0.5,1,0,0} 

\definecolor{matblue1}{rgb}{0,0.4470,0.7410}
\definecolor{matred1}{rgb}{0.85,0.325,0.098}
\definecolor{matyel1}{rgb}{0.9290, 0.6940, 0.1250}
\definecolor{matpur1}{rgb}{0.4940, 0.1840, 0.5560}
\definecolor{matgre1}{rgb}{0.4660, 0.6740, 0.1880}
\definecolor{matblue2}{rgb}{0.3010, 0.7450, 0.9330}
\definecolor{matred2}{rgb}{0.6350, 0.0780, 0.1840}
\definecolor{matgrey1}{rgb}{0.5, 0.6, 0.7}
\definecolor{matpink1}{rgb}{1, 0.07, 0.65}
\definecolor{matblue3}{rgb}{0.07, 0.62, 1}
\definecolor{gray09}{rgb}{0.9, 0.9, 0.9}

    \definecolor{mblue}{rgb}{0,0.447,0.741}
    \definecolor{mred}{rgb}{0.85,0.325,0.098}
    \definecolor{myellow}{rgb}{0.9290,0.6940,0.1250}
    \definecolor{mmagenta}{rgb}{1,0,1}
    \definecolor{mgreen}{rgb}{0.4460,0.6740,0.1880}
    \definecolor{mgrey}{rgb}{0.6,0.6,0.6}
    \definecolor{mpurple}{rgb}{0.4940, 0.1840, 0.5560}
    \definecolor{matbluevar}{rgb}{0.84314,0.89804,0.98039}%
\definecolor{matpurvar}{rgb}{0.949,0.867,1}%

    \tikzset{cross/.style={cross out, draw=black, minimum size=2*(#1-\pgflinewidth), inner sep=0pt, outer sep=0pt}, cross/.default={1pt}}
    
    \usetikzlibrary{shapes}

\newcommand{\blackdots}{\raisebox{2pt}{\tikz{\draw[-,black,densely dotted,line width = 0.9pt](0,0) -- (3mm,0);}}}

\newcommand{\bluedash}{\raisebox{2pt}{\tikz{\draw[-,matblue1,dashed,line width = 0.9pt](0,0) -- (3mm,0);}}}

\newcommand{\blueline}{\raisebox{2pt}{\tikz{\draw[-,matblue1,solid,line width = 0.9pt](0,0) -- (3mm,0);}}}
\newcommand{\redline}{\raisebox{2pt}{\tikz{\draw[-,matred1,solid,line width = 0.9pt](0,0) -- (3mm,0);}}}

\newcommand{\reddot}{\raisebox{.7pt}{\tikz{\draw[-,matred1,fill=matred1,solid,line width = 1pt](0,0) circle (.7mm);}}}

\newcommand{\yelline}{\raisebox{2pt}{\tikz{\draw[-,matyel1,solid,line width = 0.9pt](0,0) -- (3mm,0);}}}
\newcommand{\purline}{\raisebox{2pt}{\tikz{\draw[-,matpur1,solid,line width = 0.9pt](0,0) -- (3mm,0);}}}

\newcommand{\purlinedot}{\raisebox{2pt}{%
\tikz{\draw[-,matpur1,solid,line width = 0.9pt](0,0) -- (3mm,0);\draw[-,matpur1,fill=matpur1,solid,line width = 0.4mm](1.5mm,0) circle (.2mm);}}}

\newcommand{\purarea}{\raisebox{0pt}{\tikz{\draw[-,matpurvar,solid,line width = 4pt](0,0) -- (3mm,0);}}}
\newcommand{\bluearea}{\raisebox{0pt}{\tikz{\draw[-,matbluevar,solid,line width = 4pt](0,0) -- (3mm,0);}}}

\usetikzlibrary{external} 
\usepackage{graphicx}      
\usepackage{natbib}         

\usepackage{algorithm} 
\usepackage{algpseudocode}

  \usepackage{eso-pic}
  \AddToShipoutPictureBG*{%
  \AtPageUpperLeft{%
  \setlength\unitlength{1in}%
  \hspace*{\dimexpr0.5\paperwidth\relax}
  \makebox(0,-1.75)[c]{
  \begin{tabular}{c c}
  Max van Meer, 
  Cascaded Calibration of Mechatronic Systems via Bayesian Inference, \\
  Accepted for 
  {\em 22nd IFAC World Congress},
  Yokohama, Japan, 2023,
  uploaded to ArXiv \today \\
  \end{tabular}}}}

\begin{document}
\begin{frontmatter}

\title{Cascaded Calibration of Mechatronic Systems via Bayesian Inference} 

\thanks[footnoteinfo]{This work is part of the research programme VIDI with project number 15698, which is (partly) financed by the Netherlands Organisation for Scientific Research (NWO). In addition, this research has received funding from the ECSEL Joint Undertaking under grant agreement 101007311 (IMOCO4.E). The Joint Undertaking receives support from the European Union’s Horizon 2020 research and innovation programme.}

\author[TUE]{Max van Meer}
\author[TUE]{Emre Deniz}
\author[TUE,TNO]{Gert Witvoet} 
\author[TUE,Delft]{Tom Oomen}

\address[TUE]{Eindhoven University of Technology, 
   Eindhoven, the Netherlands (e-mail: m.v.meer@tue.nl).}
   \address[TNO]{TNO, Dept. of Optomechatronics, Delft, the Netherlands.}
   \address[Delft]{Delft University of Technology, Delft, the Netherlands.}

\begin{abstract}               
Sensors in high-precision mechatronic systems require accurate calibration, which is achieved using test beds that, in turn, require even more accurate calibration. The aim of this paper is to develop a cascaded calibration method for position sensors of mechatronic systems while taking into account the variance of the calibration model of the test bed. The developed calibration method employs Gaussian Process regression to obtain a model of the position-dependent sensor inaccuracies by combining prior knowledge of the sensor with data using Bayesian inference. Monte Carlo simulations show that the developed calibration approach leads to significantly higher calibration accuracy when compared to alternative regression techniques, especially when the number of available calibration points is limited. The results indicate that more accurate calibration of position sensors is possible with fewer resources.
\end{abstract}
\end{frontmatter}

\section{Introduction}
High-precision mechatronic systems rely on accurate position measurements to achieve high performance. At the same time, an increasing number of applications requires highly accurate position measurements in mass-produced systems, e.g., satellite swarms for optical communication \citep{Gregory2010,Kramer2020}, or segmented mirror telescopes \citep{Nelson2006}. 
    
The accuracy of position measurements relates to their proximity to the actual positions, and precision refers to repeatability. This paper considers sensors that consistently exhibit position-dependent inaccuracies. These repeatable sensor inaccuracies can be measured using a test bed with a more accurate sensor, e.g., coordinate measurement machines \citep{Takamasu1996} or optics-based test beds \citep{Dresscher2019}. When sensor inaccuracies are measured, a model is fitted to compensate for these inaccuracies through the process of calibration.

The test beds used to calibrate the position sensors of mechatronic systems require calibration themselves, to a standard regarded as an absolute measure of accuracy. This is done by a third party such as a metrology institute \citep{Pendrill2009}, or in-house using a highly accurate manual instrument, e.g., theodolite \citep{Krishna1996} or laser tracking interferometers \citep{Umetsu2005}. 

\begin{figure}[t]
    \centering
    \includegraphics[width=\linewidth]{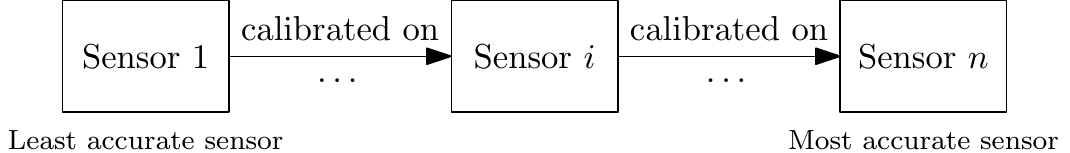}
    \caption{Schematic overview of the cascaded calibration problem. An array of sensors is calibrated on each other. Consequently, any imperfection in the calibration of the more accurate sensors is propagated down the chain to the less accurate sensors.   
    }
    \label{fig:schematic}
\end{figure}

Due to this cascade of calibration steps, depicted schematically in Fig. \ref{fig:schematic}, modeling errors in individual calibration steps can stack and limit the achieved accuracy of the sensor calibrated last. Two leading causes of modeling errors are as follows. First, calibration on manual, external calibration instruments is time-consuming. While efforts have been made to partially automate the comparison of test bed sensor readings to accurate external readings (see, e.g., \citet{Wu2013}), these methods introduce additional development cost and complexity. Consequently, the number of positions at which sensor readings are compared with those of more accurate sensors is limited. Second, some test bed locations may be unreachable to external instruments due to geometry constraints, further limiting the number of available calibration points. 

In a parallel line of developments, the application of Gaussian Process regression to mechatronic systems has gained increased attention, see \cite{Poot2022,Rasmussen2004}, since it admits a highly flexible model structure while taking uncertainty into account using Bayes' theorem. By specifying a prior that imposes properties such as smoothness and learning hyper-parameters from the data, a model is obtained that yields information not only of the expected function but also the variance of this function space. 

Although regression techniques such as lookup tables can model individual functions well if the number of calibration points is large, it is shown in this paper that by taking into account the variance of individual calibration models using Bayesian inference, a significantly more accurate model is obtained, even if the number of calibration points is limited. Moreover, while sequential calibration using Bayesian inference has attained attention in the context of computer models \citep{Jiang2020}, the literature on the cascaded calibration of position sensors is sparse.

Therefore, the aim of this paper is to find an accurate mapping of position sensor readings to `true' position values, i.e., sensor $n$ in Fig. \ref{fig:schematic}, while taking the uncertainty of the intermediate calibration model into account. The contributions of this paper are as follows:
  \begin{enumerate}[label={C\arabic*:}]
\item A method for cascaded calibration of position sensors for mechatronic systems is developed. The approach takes the model uncertainty of the first calibration step into account to arrive at a more accurate estimation in the subsequent calibration step. 
\item The effectiveness of the approach is demonstrated through Monte Carlo simulations on a reproducible case study, and it is shown that the developed calibration method yields significantly more accurate models of the sensor offsets than alternatives such as lookup tables. The results indicate that more accurate calibration of mass-produced mechatronic systems is possible with fewer resources.
\end{enumerate}
This paper is structured as follows. First, the problem description is given in Section \ref{sec:problem}. Next, the developed approach to cascaded calibration is explained in Section \ref{sec:approach}. Subsequently, simulation results are presented in Section \ref{sec:results}, and finally, conclusions are drawn in Section \ref{sec:conclusion}. 
\section{Problem description}\label{sec:problem}
In this section, the problem description is given. First, a motivating example is given. Subsequently, the calibration problem is described, and finally, the problem is formalized in terms of two regression problems.
\subsection{Motivating example}
\begin{figure}
    \centering
    \includegraphics[width=0.7\linewidth]{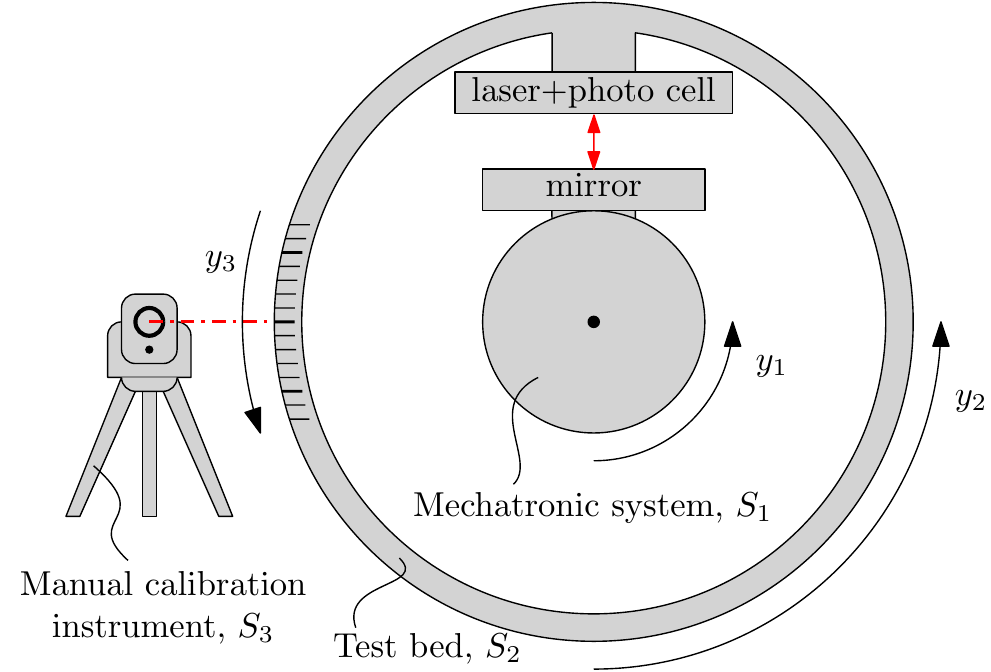}
    \caption{Motivating example. A mechatronic system with angular position sensor $S_1$ is optically linked for calibration with a test bed, with its own sensor $S_2$. The test bed itself is calibrated using a highly accurate manual measuring instrument $S_3$. Sensors $S_i$ yield different measurements $y_i$ of the same actual position when they are aligned for calibration.}
    \label{fig:motivating_ex}
\end{figure}

\begin{figure}[t]
    \centering
    \input{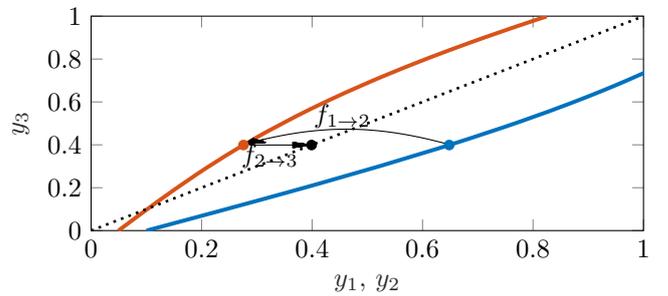}
    \caption{Sensor $S_1$ (\protect\blueline) is calibrated on a test best with sensor $S_2$ (\protect\redline), which, in turn, is calibrated on $S_3$ (\protect\blackdots). Since the systems are aligned during calibration, more accurate sensors can observe the sensor inaccuracies of less accurate sensors.}
    \label{fig:problem_exmpl}
\end{figure}

A motivating example of cascaded encoder calibration is shown in Fig. \ref{fig:motivating_ex}. The angular position sensor $S_1$ of a mechatronic system requires calibration, but the system is too compact to be accessible by the manual calibration instrument $S_3$, e.g., a theodolite or autocollimator \citep{Gao2011}. Hence, it is calibrated using a test bed with sensor $S_2$, which is optically aligned with the mechatronic system, such that readings by $S_2$ of the mechatronic system can be compared with readings by $S_1$ of the mechatronic system. The test bed, in turn, is calibrated using the manual calibration instrument, see Fig. \ref{fig:problem_exmpl}.

The manual calibration instrument $S_3$ cannot measure all locations of the test bed because its frame physically obstructs access. Moreover, accurate manual calibration is labor-intensive, especially if high accuracy is required over the entire $360^{\textnormal{o}}$ range of motion. Hence, the number of available calibration points of $S_3$ is limited. 

Any imperfection in the calibration of the test bed to the manual instrument decreases the accuracy of the mechatronic system when it is calibrated on the test bed. This propagation of modeling errors motivates the need to take the uncertainty of the calibration model of the test bed into account when calibrating the mechatronic system.
\subsection{Notation}
The following notation is used. Sensor $S_i$, $i\in\{1,2,3\}$, is fixed to system $i$, where system 1 is the mechatronic system, system 2 is the test bed, and system 3 is the manual calibration instrument. Sensor $S_i$ yields measurement ${y}_i\in\mathbb{R}$. All positions are defined w.r.t. the same fixed reference frame. The true position of system $i$ is denoted by $y_i^{*}$. When systems $i$ and $j$ are aligned for calibration (as detailed in Assumption \ref{ass:align}), it holds that $y_i^{*}=y_j^*$, and hence, measurements ${y}_i$ and ${y}_j$ are two different measurements of the same true position. 

\subsection{Cascaded calibration of sensors}
Sensor $S_i$ is generally not perfectly accurate, i.e., ${y}_i\neq y_i^*$. Sensor inaccuracies can have multiple causes, e.g., manufacturing tolerances, mechanical wear, or imperfect assembly. These inaccuracies to the true position are position-dependent, expressed as follows: \begin{equation}
y_i^* = f_{i\rightarrow i^*}({y}_i),
\end{equation}
where $f_{i\rightarrow i^*}:\mathbb{R}\to\mathbb{R}$ is a possibly nonlinear function, mapping inaccurate position measurements to true positions.
\begin{assumption}
    Measurements ${y}_i$ of $y_i^*$ are corrupted by zero-mean additive Gaussian white noise $\varepsilon_i$ with variance $\sigma_{n,i}^2$ assumed small compared to the sensor inaccuracies. 
    Long-term temporal changes in sensor-inaccuracies are assumed negligible, and short-term changes are assumed zero-mean, such that temporal effects are averaged out over multiple observations at the same location, i.e., $f_{i\rightarrow i^*}({y}_i,t)\approx f_{i\rightarrow {i^*}}({y}_i)$.
\end{assumption}
If sensor $S_i$ of system $i$ is not perfectly accurate, then these inaccuracies are measured by a second system $j$, provided that sensor $S_j$ is more accurate, i.e., \begin{equation}
\int_{y_{j\min}}^{y_{j,\max}}|{y}_j(y_j^*)-y_j^* | dy_j^* < \int_{y_{i\min}}^{y_{i,\max}}|{y}_i(y_i^*)-y_i^* |dy_{i}^*.
\end{equation}
System $j$ can only be used to measure the sensor inaccuracies of $S_i$ if systems $i$ and $j$ are aligned, i.e., $y_i^*=y_j^*$, because only then do they attempt to measure the same true position. Therefore, the following assumption is made:\begin{assumption}\label{ass:align}
When a pair $({y}_i,{y}_{j})$ of sensor readings is measured at a fixed point in time, it is assumed that misalignment errors are negligible w.r.t. sensor inaccuracies:
\begin{equation}\label{eq:alignment}
|y_{i}^* - y_{j}^* | \ll |y_{i}^*-{y}_i|.
\end{equation}
Hence, both measurements $({y}_i,{y}_{j})$ are assumed to describe the same true position $y_i^*\approx y_j^*$ during calibration.
\end{assumption}

The manual calibration instrument $S_3$ is the most accurate sensor available, and therefore, ${y}_3$ can effectively be used as a proxy for $y_3^*$. 
The following function is then defined, assuming systems 3 and $i$ are aligned:
\begin{equation}
y_3^* = f_{i\rightarrow 3}({y}_i),
\end{equation}
i.e., $f_{i\rightarrow 3}$ describes the relationship between a sensor reading ${y}_i$ and its `true' position $y_3^*\approx {y}_3$. 
\begin{assumption}\label{ass:function}
    Each $f_{i\rightarrow 3}({y}_i)$, $i
    \in\{1,2\}$, is bijective, i.e., any value of ${y}_i$ maps to one value of $y_3^*$ and vice versa. \end{assumption}
The aim is to obtain a model $\hat{f}_{1\rightarrow 3}$ of ${f}_{1\rightarrow 3}$, describing the sensor inaccuracy of $S_1$, but importantly, ${f}_{1\rightarrow 3}$ is never observed directly. In other words, the manual calibration instrument is not aligned with the mechatronic system for calibration, for two reasons: $(i)$ it is economically not viable to perform manual calibration on many different mechatronic systems with sensor $S_1$, and $(ii)$ the mechatronic system may be built too compactly to be physically accessible by a manual calibration instrument.

To prevent the need to calibrate $S_1$ on $S_3$ directly, the test bed with $S_2$ is first calibrated on $S_3$. Subsequently, $S_2$ can calibrate many different mechatronic systems, each with its own sensor $S_1$. These two steps are detailed in Procedure \ref{proc:casc}. When $\hat{f}_{1\rightarrow 3}$ is modeled offline through Procedure \ref{proc:casc}, it is used online to transform inaccurate position measurements $y_1$ to corrected measurements $\hat{f}_{1\rightarrow 3}(y_1)$.

\makeatletter
\renewcommand*{\ALG@name}{Procedure}
\makeatother
\begin{algorithm}[t]
\caption{Cascaded calibration of $S_1$ on $S_3$ via $S_2$}\label{proc:casc}
\begin{algorithmic}[1]
\State Align systems 2 and 3 and obtain a data-set $\mathcal{D}_2 =\{\bar{y}_{2,k}, \bar{y}_{3,k}\}_{k=1}^{N_2}$. Use these observations of $\bar{y}_3=f_{2\rightarrow 3}(\bar{y}_2)$ to fit a function $y_3=\hat{f}_{2\rightarrow 3}(y_2)$.
\State Align systems 1 and 2 and obtain a data-set $\mathcal{D}_1 = \{\bar{y}_{\underline{1},k}, \bar{y}_{\underline{2},k}\}_{k=1}^{N_1}$. Then construct $\mathcal{D}_1^{\prime}=\{\bar{y}_{\underline{1},k}, \hat{f}_{2\rightarrow 3}(\bar{y}_{\underline{2},k})\}_{k=1}^{N_1}$ using the model of Step 1, and use these `observations' of $\hat{f}_{2\rightarrow 3}(\bar{y}_{\underline{2}})=f_{1\rightarrow 3}(\bar{y}_1)$ to fit a function $y_3=\hat{f}_{1\rightarrow 3}(y_1)$.
    \end{algorithmic}
\end{algorithm}
Importantly, cascaded calibration requires making a fit on another fit. Since measurements of $S_3$ are labor-intensive and time-consuming, the first model $\hat{f}_{2\rightarrow 3}$ might be based on a limited amount of data ($N_2 \ll N_1$), and consequently, it may have a large variance. This potentially deteriorates the accuracy of $\hat{f}_{1\rightarrow 3}$ w.r.t. the true ${f}_{1\rightarrow 3}$. 

In the next section, it is explained how the construction of $\hat{f}_{1\rightarrow 3}(y_1)$ from data through Steps 1-2 of Procedure \ref{proc:casc} is framed as a series of regression problems.
\subsection{Cascaded calibration through regression}
To obtain a model $\hat{f}_{1\rightarrow 3}$ of the sensor inaccuracies of $S_1$ that can be used for calibration, the following cascade of regression problems is defined.


\begin{problem}\label{prob:1}
Consider Procedure \ref{proc:casc}, and suppose that data-sets $\mathcal{D}_1$ and $\mathcal{D}_2$ are available. Parameterize the models 
as $y_3=\hat{f}_{1\rightarrow 3}(y_1,\alpha)$ and $y_3=\hat{f}_{2\rightarrow 3}(y_2,\beta)$, respectively, and let their structures be fully determined by Hilbert spaces $\mathcal{K}_{1}$ and $\mathcal{K}_2$ \citep{Wegman2006}.
The aim is to obtain the best possible fit of $f_{1\rightarrow 3}$, even though $f_{1\rightarrow 3}$ is not measured directly, but instead by solving two sub-problems:
\begin{equation}\label{eq:problem}
    \begin{aligned}
\min_{\mathcal{K}_1,\mathcal{K}_2}\ \mathcal{J} = &
\left(\frac{\int_{y_{1}^{\textnormal{min}}}^{y_{1}^{\textnormal{max}}} [\hat{f}_{1\rightarrow 3}\left({y}_{{1}},\alpha^\star\right) - f_{1\rightarrow 3}({y}_{{1}})]^2 \d y_1}{y_{1}^{\textnormal{max}}-y_{1}^{\textnormal{min}}}\right)^{\frac{1}{2}}\\
\text{subject to}&\\
  \beta^\star = {\text{argmin}} &\left\|\hat{f}_{2\rightarrow 3}\left(\bar{y}_2,\beta\right)-{f}_{2\rightarrow 3}\left(\bar{y}_2\right)\right\|_2^2 + \left\|\hat{f}_{2\rightarrow 3}\right\|_{\mathcal{K}_2}^2,\\
   \alpha^\star ={\text{argmin}} &\left\|\hat{f}_{1\rightarrow 3}\left(\bar{y}_{\underline{1}},\alpha\right)- \hat{f}_{2\rightarrow 3}(\bar{y}_{\underline{2}},\beta^\star)    \right\|_2^2+ \left\|\hat{f}_{1\rightarrow 3}\right\|_{\mathcal{K}_1}^2,
\end{aligned}
\end{equation}
where $y_1^{\textnormal{min}}$ and $y_1^{\textnormal{max}}$ specify a range of positions where a good model of the sensor inaccuracies of $S_1$ is desired.  
Note that solving for $\beta^{\star}$ and for $\alpha^{\star}$ amounts to constructing the fits of Steps 1 and 2 in Procedure \ref{proc:casc}, respectively. 
\end{problem}
The cost $\mathcal{J}$ cannot be evaluated in practice because $S_1$ can only be compared to $S_2$, and $S_2$ to $S_3$, but not $S_1$ to $S_3$. On the other hand, $\mathcal{J}$ can be evaluated in simulation, when $f_{1\rightarrow 3}$ is known. In the following sections, it is shown that for a specific choice of the structure of $\hat{f}$ through $\mathcal{K}_1$ and $\mathcal{K}_2$, the cost $\mathcal{J}$ is significantly reduced, when compared to conventional regression methods, indicating that more accurate calibration is achievable. 
\section{Cascaded calibration via Bayesian inference}\label{sec:approach}
In this section, the developed solution to Problem \ref{prob:1} is explained. The key idea is to recognize that the two sub-problems in \eqref{eq:problem} need to be posed in a coupled fashion. If these problems were solved independently, then any inaccuracy in $\hat{f}_{2\rightarrow 3}(y_2)$ that follows from the fact that $y_3$ is only available at a limited number of positions is over-confidently carried over to $\hat{f}_{1\rightarrow 2}(y_1)$. 

Instead, the developed approach takes into account the uncertainty of $\hat{f}_{2\rightarrow 3}(y_2)$ at locations $y_2$ where no data of $y_3$ is available, through Bayes' rule. 

\subsection{Calibration of $S_2$}\label{sec:s2}
First, sensor $S_2$ needs to be calibrated on sensor $S_3$; see the first sub-problem in \eqref{eq:problem}. Given a limited number of observations of pairs $({y}_2,{y}_3)$ showing the relation ${y}_3 = f_{2\rightarrow 3}({y}_2)$, there is uncertainty in $\hat{f}_{2\rightarrow 3}({y}_2,\beta)$ for values of $y_2$ that are far from calibrated locations. It is explained next how this model uncertainty, or variance, is computed explicitly so that it can be used for more accurate regression in the next section. To this end, a probabilistic viewpoint is adopted. 

The model structure of $\hat{f}_{2\rightarrow 3}(y_2,\beta)$ is assumed to be \begin{equation}
\hat{f}_{2\rightarrow 3}(y_2,\beta) = \boldsymbol{\phi}_2(y_2)^\top \beta,
\end{equation}
where $\boldsymbol{\phi}_2: \mathbb{R}\to \mathbb{R}^{D}$ maps any $y_2$ into some $D$-dimensional feature space, with weights $\beta$.
A Gaussian prior is assumed on $\beta$, i.e., \begin{equation}
p(\beta) = \mathcal{N}\left(\beta_0, \Sigma_{2,p}\right),
\end{equation} 
with mean $\beta_0$ and prior variance $\Sigma_{2,p} \in \mathbb{R}^{D\times D}$. With this model, the likelihood $p(\bar{Y}_3\mid \bar{Y}_2,\beta)$, or the probability density of the observations given the parameters, is given by \begin{equation}
p(\bar{Y}_3\mid \bar{Y}_2,\beta) = \mathcal{N}({\Phi}_2(\bar{Y}_2)^\top \beta, \Sigma_{\bar{Y}_3}),
\end{equation}
where \begin{equation}
\begin{aligned}
\bar{Y}_i&:=[\bar{y}_{i,1},\ldots,\bar{y}_{i,N_i}]^\top,\quad & i\in\{1,2\},\\
{\Phi}_i(\bar{Y}_i)&:= [\boldsymbol{\phi}_i(\bar{y}_{i,1}),\ldots,\boldsymbol{\phi}_i(\bar{y}_{i,N_i})]^\top, \quad & i\in\{1,2\},
\end{aligned}
\end{equation}
and the variance of the observations is given by
\begin{equation}
\Sigma_{\bar{Y}_3} = \sigma_{n,3}^2 I.
\end{equation} 
The prior on $\beta$ is conditioned on the data $\mathcal{D}_2$ to obtain the posterior distribution, i.e., the probability of the parameters given the data. From Bayes' rule, it is known that \begin{equation}
\text{posterior} = \frac{\text{likelihood}\times\text{prior}}{\text{marginal likelihood}},
\end{equation}
or specifically,
\begin{equation}\label{eq:posterior_beta}
p(\beta\mid \bar{Y}_3, \bar{Y_2}) = \frac{p(\bar{Y}_3\mid \bar{Y}_2,\beta) p(\beta)}{p(\bar{Y}_3\mid\bar{Y}_2)}.
\end{equation} 

The expression \eqref{eq:posterior_beta} describes the posterior distribution of the parameters. For regression, rather, the predictive distribution  $p({Y}_3 \mid \bar{Y}_2,\bar{Y}_3)=p(\hat{f}_{2\rightarrow 3}(Y_2)\mid \bar{Y}_2,\bar{Y}_3)$ is of interest. For arbitrary sensor positions $Y_2\in\mathbb{R}^{M_2}$, this predictive distribution is computed as \begin{equation}
p(\hat{f}_{2\rightarrow 3}(Y_2)\mid \bar{Y}_2,\bar{Y}_3) = \int p({Y}_3\mid {Y}_2,\beta) p(\beta\mid \bar{Y}_2,\bar{Y}_3) d\beta, 
\end{equation}
which is again a Gaussian with mean and variance
\begin{equation}\label{eq:posterior_2}
\begin{aligned}
\mathbb{E}\left[\hat{f}_{2\rightarrow 3}(Y_2)\right] =&m_2(Y_2)+K_{2}(Y_2,\bar{Y}_2)\cdot\\
&\hspace{0cm}\left[K_2(\bar{Y}_2,\bar{Y}_2)+\Sigma_{\bar{Y}_3}\right]^{-1} (\bar{Y}_3-m_2(\bar{Y}_2)),\\
\textnormal{cov}(\hat{f}_{2\rightarrow 3}(Y_2)) =& K_{2}(Y_2,{Y}_2)-K_{2}(Y_2,\bar{Y}_2)\cdot\\
&[K_{2}(\bar{Y}_2,\bar{Y}_2)+\Sigma_{\bar{Y}_3}]^{-1} K_{2}(\bar{Y}_2,{Y}_2),
\end{aligned}
\end{equation}

and the elements of $K_i(\bar{Y}_i,Y_i)=K_i({Y}_i,\bar{Y}_i)^\top\in\mathbb{R}^{N_i\times M_i}$, $K_i(\bar{Y}_i,\bar{Y}_i)\in\mathbb{R}^{N_i\times N_i}$ and $K_i({Y}_i,Y_i)\in\mathbb{R}^{M_i\times M_i}$ are obtained from evaluating a kernel function $k_i(y_A,y_B)$ for the corresponding values of arbitrary positions $Y_i$ and measurements $\bar{Y}_i$. This kernel has the property that $K_i = \Phi_i^\top \Sigma_{i,p} \Phi_i$, and thus relates to the chosen model structure or prior. The prior mean is denoted by $m_i:\mathbb{R}^{M_i}\to \mathbb{R}^{M_i}$. See Section \ref{sec:kernel} for details on the choice of $k_i$ and $m_i$.

\begin{figure*}[t]
\hspace{-0.4cm}\input{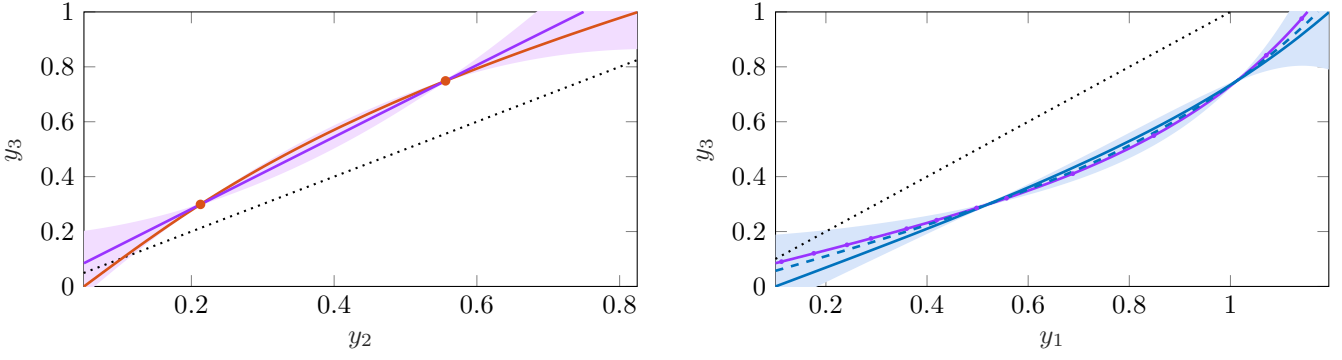}
\caption{Example of cascaded calibration via Bayesian inference. Sensor readings $y_2$ (\protect\redline) and $y_1$ (\protect\blueline) have different position-dependent inaccuracies w.r.t. $y_3\approx y^*$ (\protect\blackdots) while the systems are physically aligned to $y^*$. $\bar{y}_3$ is measured at two locations (\protect\reddot) and a model $\hat{f}_{2\rightarrow 3}(y_2)$ (\protect\purline,\protect\purarea) is fitted through these points. Next, this model is used to obtain $\hat{{y}}_3=\hat{f}_{2\rightarrow 3}(\bar{y}_{\underline{2}})$  (\protect\purlinedot,right) for many measurements $\bar{y}_2$. Finally, a model $\hat{f}_{1\rightarrow 3}$ (\protect\bluedash,\protect\bluearea) is fitted on $\hat{{y}}_3$. Because this model takes the uncertainty of $\hat{f}_{2\rightarrow 3}$ into account, the model $\hat{f}_{1\rightarrow 3}$ (\protect\bluedash) partially disregards the data at locations with high uncertainty, and instead relies more on its prior $\mathbb{E}[y_3]=y_1$, leading to a more accurate fit to $f_{{1\rightarrow 3}}$ (\protect\blueline).}
\label{fig:uncertain_sketch}
\end{figure*}

Crucially, \eqref{eq:posterior_2} provides an analytic expression of the covariance of $\hat{f}_{2\rightarrow 3}(Y_2)$, i.e., the uncertainty of the model, illustrated in light purple in the left of Fig. \ref{fig:uncertain_sketch}. This covariance is instrumental to obtaining a more accurate estimate $\hat{f}_{1\rightarrow 3}$, as explained in the next section.

\subsection{Calibration of $S_1$}
In this section, a model $\hat{f}_{1\rightarrow 3}$ is made, based on the model of $\hat{f}_{2\rightarrow 3}$ obtained in the previous section. Since sensor readings $S_1$ can only be experimentally compared with $S_2$, but not to $S_3$, the model of $\hat{f}_{2\rightarrow 3}$ is applied to $\mathcal{D}_1 = \{\bar{y}_{{\underline{1}},k}, \bar{y}_{\underline{2},k}\}_{k=1}^{N_1}$ to obtain \begin{equation} \label{eq:Dprime}
\begin{aligned}
\mathcal{D}_1^\prime =& \{\bar{y}_{{\underline{1}},k}, \hat{{{y}}}_{3,k}\}_{k=1}^{N_1},
\end{aligned}
\end{equation}
where \begin{equation}
\begin{aligned}
\hat{{{y}}}_{3,n}&:=\hat{f}_{2\rightarrow 3}(\bar{y}_{\underline{2},n}),\\
\hat{{Y}}_3 &:= [\hat{f}_{2\rightarrow 3}(\bar{y}_{\underline{2},1}),\ldots,\hat{f}_{2\rightarrow 3}(\bar{y}_{\underline{2},N_1})]^\top.
\end{aligned}
\end{equation}

The key insight that distinguishes the regression approach in this paper from traditional methods is that the variance of the `observations' $\hat{{y}}_3=\hat{f}_{2\rightarrow 3}(\bar{y}_2)\in\mathcal{D}_1^\prime$, i.e., the prediction of the model created in the previous section evaluated at measurements $\bar{y}_2$, is affected by the uncertainty of the model. The covariance matrix corresponding to $\hat{Y}$ follows directly from this model uncertainty and is given by
\begin{equation}
\Sigma_{\hat{{Y}}_3}:= \text{cov}(\hat{f}_{2\rightarrow 3}(\bar{Y}_{\underline{2}})),
\end{equation}
which is computed directly through \eqref{eq:posterior_2}. By assuming a Gaussian prior on $\hat{f}_{1\rightarrow 3}$ as before and conditioning on $\mathcal{D}_1^\prime$, the predictive distribution $p(\hat{f}_{1\rightarrow 3})$ is a Gaussian with \begin{equation}\label{eq:posterior_1}
\begin{aligned}
\mathbb{E}\left[\hat{f}_{1\rightarrow 3}(Y_1)\right] =&m_1(Y_1) + K_{1}(Y_1,\bar{Y}_{\underline{1}})^\top\cdot\\
&\left[K_1(\bar{Y}_{\underline{1}},\bar{Y}_{\underline{1}})+\Sigma_{\hat{{Y}}_3}\right]^{-1} (\hat{{Y}}_3-m_1(\bar{Y}_{\underline{1}})),\\
\textnormal{cov}(\hat{f}_{1\rightarrow 3}(Y_1)) =& K_{1}(Y_1,{Y}_1)-K_{1}(Y_1,\bar{Y}_{\underline{1}})\cdot\\
&[K_{1}(\bar{Y}_{\underline{1}},\bar{Y}_{\underline{1}})+\Sigma_{\hat{{Y}}_3}]^{-1} K_{1}(\bar{Y}_{\underline{1}},{Y}_1).
\end{aligned}
\end{equation}
Indeed, the posterior mean $\mathbb{E}[\hat{f}_{1\rightarrow 3}]$, shown in dashed blue in Fig. \ref{fig:uncertain_sketch}, is a function of $\Sigma_{\hat{{Y}}_3} = \text{cov}(\hat{f}_{2\rightarrow 3})$, shown in light purple. Clearly, by taking into account the uncertainty of the model $\hat{f}_{2\rightarrow 3}$, the model $\hat{f}_{1\rightarrow 3}$ is affected. It is shown in Section \ref{sec:results} that this choice results in a more accurate model $\hat{f}_{1\rightarrow 3}$ than when the uncertainty is ignored. This concept is shown visually in Fig. \ref{fig:uncertain_sketch}, where the key observation is that $\hat{f}_{1\rightarrow 3}$ (dashed blue) does not rely on $\hat{y}_3$ (purple dots) at locations where $\hat{f}_{2\rightarrow 3}$ has high variance (light purple), but instead relies more on its prior $\mathbb{E}[y_3]=y_1$. 

Next, the choice of $k_i$ and $m_i$, which determine the prior of $\hat{f}_{i\rightarrow j}$, is explained.
\subsection{Selection of the model structure and hyper-parameters}\label{sec:kernel}
In this section, the chosen model structure, or prior, is elaborated, and it is shown how the hyper-parameters are chosen automatically using data with empirical Bayes. 

First, the choice of the prior is explained. It follows from Equations \eqref{eq:posterior_2}, \eqref{eq:posterior_1} that the model structure of $\hat{f}_{i\rightarrow j}$ is uniquely determined by choice of kernel function $k_i$, defining the prior variance, and the prior mean $m_i$. 

For a function $y_j=\hat{f}_{i\rightarrow j}(y_i)$, the prior mean is \begin{equation}
m_i(y_i) := \mathbb{E}[y_j].
\end{equation}
An intuitive choice is to pick $m_i(y_i) = y_i$, which assumes that in the absence of observations, sensor $S_j$ is expected to yield identical measurements as $S_i$. If prior information on the sensor inaccuracy is available, it can also be incorporated into $m_i$. 

The prior variance, i.e., the range of possible functions that $\hat{f}_{i\rightarrow j}$ can take, is determined by the kernel function. For an overview of possible kernel functions, including, e.g., polynomial or sinusoidal model structures, see \citet{Duvenaud2014}. In this paper, the attention is restricted to squared exponential (SE) kernel functions of the form \begin{equation}\label{eq:se_kernel}
k_i(y_A,y_B) = \sigma_{f,i}^2 \exp\left(-\frac{1}{2 \ell_i^2}(y_A-y_B)^2 \right),
\end{equation}
where hyper-parameters $\ell_i$ and $\sigma_{f,i}^2$ are the characteristic length scale and the magnitude of the prior variance, respectively. This model structure can be interpreted as imposing smoothness on $\hat{f}_{i\rightarrow j}$. 

Hyper-parameters $\Theta_i=\{\ell_i$, $\sigma_{f,i}^2$, $\sigma_{n,i}^2\}$ can be chosen from prior knowledge of the smoothness and magnitude of the sensor inaccuracies and noise. Alternatively, these can be learned from the data $\mathcal{D}_i$, also known as empirical Bayes, by maximizing the log marginal likelihood, given by \begin{equation}\label{eq:marginal}
\begin{aligned}
\text{log}\ p(Y_{j} | Y_i, \Theta_i) =& -\frac{1}{2} Y_{j}^\top \tilde{K}_i^{-1} Y_{j} \\
&- \frac{1}{2} \text{log} \left| \tilde{K}_i \right|-\frac{N_i}{2} \text{log} 2\pi,
\end{aligned}
\end{equation}
with $\tilde{K}_i= K_i(Y_{i},Y_{i})+\Sigma_{Y_j}$. This expression is maximized with respect to $\Theta$ using an optimization algorithm for non-convex problems, to find optimal hyper-parameters.
\subsection{Summary}
The complete algorithm to obtain estimates of $y_3$ for arbitrary measurements $y_1$ is summarized in Algorithm \ref{alg:bayesian}. Note that after following steps 1-5, step 6 can be repeated cheaply for any $Y_1$, since the computation of $\mathbb{E}[Y_3] = \mathbb{E}[\hat{f}_{1\rightarrow 3}(Y_{{1}})]$ through \eqref{eq:posterior_1} is simply a matrix-vector multiplication once the inverted matrix in \eqref{eq:posterior_1} is stored for future use. 
\begin{remark}
The posterior means in \eqref{eq:posterior_2} and \eqref{eq:posterior_1} are identical to the solutions of the sub-problems in \eqref{eq:problem}, see \citet[Section 6.2]{Rasmussen2004} for details.
\end{remark}

\setcounter{algorithm}{0}
\makeatletter
\renewcommand*{\ALG@name}{Algorithm}
\makeatother
\begin{algorithm}[t]
	\caption{Cascaded calibration via Bayesian inference}\label{alg:bayesian}
	\begin{algorithmic}[1]
	\Require Data-sets $\mathcal{D}_1$, $\mathcal{D}_2$, test points $Y_1\in\mathbb{R}^{M_1}$.
	\State Specify kernel functions $k_1$, $k_2$ with initial hyper-parameters $\Theta_{i,0}$, see Section \eqref{sec:kernel}.
	\State Find optimal hyper-parameters $\Theta_2^\star$ by maximization of $\text{log}\ p(\bar{Y}_{3} | \bar{Y}_2, \Theta_2)$, see \eqref{eq:marginal}.
	\State Compute $\mathbb{E}[\hat{{Y}}_3] = \mathbb{E}[\hat{f}_{2\rightarrow 3}(\bar{Y}_{\underline{2}})]\in\mathbb{R}^{N_1}$ and $\text{cov}(\hat{{Y}}_3)=\text{cov}(\hat{f}_{2\rightarrow 3}(\bar{Y}_{\underline{2}}))\in\mathbb{R}^{N_1\times N_1}$ with \eqref{eq:posterior_2}.
	\State Find optimal hyper-parameters $\Theta_1^\star$ by maximization of $\text{log}\ p(\hat{{Y}}_{3} | \bar{Y}_{\underline{1}}, \Theta_1)$, see \eqref{eq:marginal}.
	\State Compute $\mathbb{E}[Y_3] = \mathbb{E}[\hat{f}_{1\rightarrow 3}(Y_{{1}})]\in\mathbb{R}^{M_1}$ with \eqref{eq:posterior_1}. \\

    \Return $\mathbb{E}[Y_3]$
	\end{algorithmic} 
\end{algorithm}
\section{Results}\label{sec:results}
In this section, the effectiveness of the developed calibration approach is demonstrated through Monte Carlo simulations. The simulation set-up is given first, and subsequently, the results are presented.
\subsection{Monte Carlo simulation set-up}
Suppose the sensor measurements ${y}_i$ obtained from sensor $S_i$ during alignment ($y_1^*=y_2^*=y_3^*=y^*$) are described by \begin{equation}\label{eq:modelstructure}
\begin{aligned}
{y}_3 &= y^* + \varepsilon, \\
{y}_2 &= y^* +\sum_{k=1}^{N_s} a_k\sin\left(\omega_{1,k} y^*\right) + b_k\cos\left(\omega_{1,k} y^*\right) + \varepsilon,\\
{y}_1 &= y^* +\sum_{k=1}^{N_s} c_k\sin\left(\omega_{2,k} y^*\right)+d_k\cos\left(\omega_{2,k} y^*\right) + \varepsilon,
\end{aligned}
\end{equation}
with $N_s$ = 10 and $\varepsilon\sim\mathcal{N}(0,10^{-8})$. The range of interest is $y^*\in[0, 1]$ m, and hence, the values $y_1^{\textnormal{min}}$ and $y_1^{\textnormal{max}}$ in \eqref{eq:problem} follow from \eqref{eq:modelstructure}. 

For the Monte Carlo simulations, $N=12000$ different pairs of functions \eqref{eq:modelstructure} are generated with $a_k, b_k, c_k, d_k \sim \mathcal{N}(0,10^{-4})$ and $\omega_{i,k}\sim\mathcal{N}(0,6)$. 
The data sets are collected as follows. Data-set $\mathcal{D}_1$ is obtained by observing ${y}_{{2}}$ for an equally spaced grid of $N_1=100$ values of ${y}_{{1}}$. Subsequently, ${y}_2$ is observed for an equally spaced grid of $100$ values of ${y}_3$, but then 10\% of the data on either edge and 20\% of the data in the center is removed, leading to $N_2=64$. This represents a scenario where $S_3$ cannot measure the test bed everywhere because it is physically obstructed. 

Algorithm \ref{alg:bayesian} is followed for all $N$ functions. The accuracy of the resulting model $\hat{f}_{1\rightarrow 3}$ is then assessed by its cost $\mathcal{J}$ in \eqref{eq:problem} and compared with two alternative techniques:
\begin{enumerate}[label={Alternative $\arabic*$:},align=left]
    \item Algorithm \ref{alg:bayesian} is followed, except that uncertainty of the model $\hat{f}_{2\rightarrow 3}$ is approximated as $\Sigma_{\hat{{Y}}_3}=\sigma_{n,3}^2 I$, where $\sigma_{n,3}^2$ is found by maximization of \eqref{eq:marginal}. 
    \item A lookup table is made of $\hat{f}_{2\rightarrow 3}$ using $\mathcal{D}_2$. Subsequently, a lookup table $\hat{f}_{1\rightarrow 3}$ is made using $\mathcal{D}_1^\prime$, see \eqref{eq:Dprime}. Linear interpolation is used between entries in the lookup table.
\end{enumerate}
\begin{figure}[t]
    \centering
\input{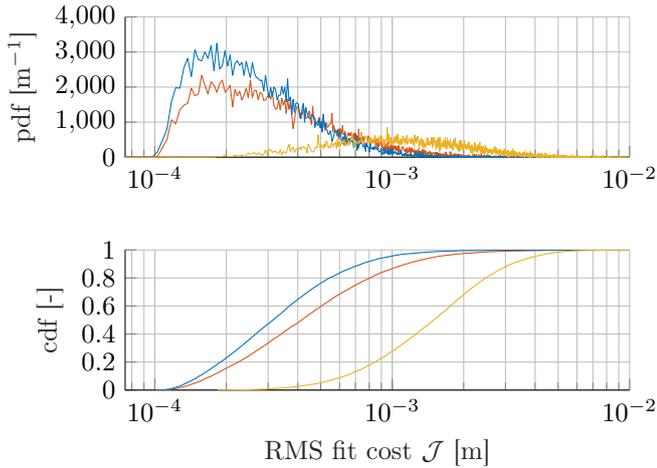}

\caption{Normalized empirical probability density functions (top) and cumulative density functions (bottom) of $\mathcal{J}$ (N=12000). Algorithm \ref{alg:bayesian} (\protect\blueline) leads to a more accurate fit $\hat{f}_{1\rightarrow 3}$ than Alternative 1 (\protect\redline), in which the variance of the first model is ignored. Both methods perform significantly better than Alternative 2 (\protect\yelline), a lookup table with linear interpolation.}    \label{fig:case2_results}
\end{figure}
\subsection{Simulation results}
The computational times of Algorithm \ref{alg:bayesian} and Alternative 1 were less than five seconds each on a personal computer for each of the $N=12000$ simulations. The results of the Monte Carlo simulations are shown in Fig. \ref{fig:case2_results}. The empirical probability distribution functions are normalized to have area 1. It is clear from the Fig. \ref{fig:case2_results} that taking the model uncertainty of $\hat{f}_{2\rightarrow 3}$ into account through Algorithm \ref{alg:bayesian} leads to a considerably better fit than when the uncertainty is ignored (Alternative 1). Both kernel-based methods result in more accurate models than a lookup table with linear interpolation (Alternative 2).

The results indicate that Algorithm \ref{alg:bayesian} is a suitable solution to Problem \ref{eq:problem}, leading to better models of sensor inaccuracies than alternative approaches. With better models of sensor inaccuracies, more accurate calibration is achieved. 





\section{Conclusion and recommendations}\label{sec:conclusion}
A cascaded calibration method is developed to accurately model position-dependent inaccuracies of position sensors, enabling more accurate calibration of mass-produced systems in less time. By taking into account the uncertainty resulting from the first regression step using Bayesian inference, more accurate calibration is achieved than conventional methods such as lookup tables. The approach is especially advantageous when the number of calibration points is limited. Moreover, since the model hyper-parameters are tuned automatically using the data, the procedure is convenient to implement in practice.

Future work is required to deal with cases when sensor readings are not fully repeatable. In the current framework, this is done by storing only average readings in the data sets, but a proper Bayesian treatment of this spread might further increase the achieved accuracy. Moreover, future efforts will be aimed at experimental validation of the method on the motivating example. 
\bibliography{library.bib}

\begin{thebibliography}{15}
\providecommand{\natexlab}[1]{#1}
\providecommand{\url}[1]{\texttt{#1}}
\providecommand{\urlprefix}{URL }
\expandafter\ifx\csname urlstyle\endcsname\relax
  \providecommand{\doi}[1]{doi:\discretionary{}{}{}#1}\else
  \providecommand{\doi}{doi:\discretionary{}{}{}\begingroup
  \urlstyle{rm}\Url}\fi

\bibitem[{Dresscher et~al.(2019)Dresscher, Human, Witvoet, {Van Der Heiden},
  {Den Breeje}, Kuiper, Fritz, Korevaar, {Van Der Valk}, {De Lange}, Saathof,
  Doelman, Crowcombe, Duque, and {De Man}}]{Dresscher2019}
Dresscher, M., Human, J.D., Witvoet, G., {Van Der Heiden}, N., {Den Breeje},
  R., Kuiper, S., Fritz, E.C., Korevaar, C.W., {Van Der Valk}, N.C., {De
  Lange}, T.J., Saathof, R., Doelman, N., Crowcombe, W.E., Duque, C.M., and {De
  Man}, H. (2019).
\newblock {Key Challenges and Results in the Design of Cubesat Laser Terminals,
  Optical Heads and Coarse Pointing Assemblies}.
\newblock \emph{2019 IEEE Int. Conf. on Space Optical Systems and Applications,
  ICSOS 2019}.

\bibitem[{Duvenaud(2014)}]{Duvenaud2014}
Duvenaud, D. (2014).
\newblock \emph{{Automatic model construction with Gaussian processes}}.
\newblock Ph.D. thesis, University of Cambridge.

\bibitem[{Gao et~al.(2011)Gao, Saito, Muto, Arai, and Shimizu}]{Gao2011}
Gao, W., Saito, Y., Muto, H., Arai, Y., and Shimizu, Y. (2011).
\newblock {A three-axis autocollimator for detection of angular error motions
  of a precision stage}.
\newblock \emph{CIRP Annals - Manufacturing Technology}, 60(1), 515--518.

\bibitem[{Gregory et~al.(2010)Gregory, Heine, K{\"{a}}mpfner, Meyer, Fields,
  and Lunde}]{Gregory2010}
Gregory, M., Heine, F., K{\"{a}}mpfner, H., Meyer, R., Fields, R., and Lunde,
  C. (2010).
\newblock {TESAT laser communication terminal performance results on 5.6Gbit
  coherent inter satellite and satellite to ground links}.
\newblock In N.~Kadowaki (ed.), \emph{International Conference on Space Optics
  — ICSO 2010}, vol. 10565, 37. SPIE.

\bibitem[{Jiang et~al.(2020)Jiang, Hu, Liu, Mourelatos, Gorsich, and
  Jayakumar}]{Jiang2020}
Jiang, C., Hu, Z., Liu, Y., Mourelatos, Z.P., Gorsich, D., and Jayakumar, P.
  (2020).
\newblock {A sequential calibration and validation framework for model
  uncertainty quantification and reduction}.
\newblock \emph{Computer Methods in Applied Mechanics and Engineering}, 368,
  113172.

\bibitem[{Kramer et~al.(2020)Kramer, Peters, Voorhoeve, Witvoet, and
  Kuiper}]{Kramer2020}
Kramer, L., Peters, J., Voorhoeve, R., Witvoet, G., and Kuiper, S. (2020).
\newblock {Novel motorization axis for a Coarse Pointing Assembly in Optical
  Communication Systems}.
\newblock In \emph{IFAC PapersOnLine}, vol.~53, 8426--8431. Elsevier Ltd.

\bibitem[{Krishna(1996)}]{Krishna1996}
Krishna, R. (1996).
\newblock {Improved pointing accuracy using high-precision theodolite
  measurements}.
\newblock In E.R. Washwell (ed.), \emph{GOES-8 and Beyond}, vol. 2812, 199 --
  209. International Society for Optics and Photonics, SPIE.

\bibitem[{Nelson(2006)}]{Nelson2006}
Nelson, J. (2006).
\newblock Segmented mirror telescopes.
\newblock In \emph{Optics in Astrophysics}, 61--72. Springer.

\bibitem[{Pendrill(2009)}]{Pendrill2009}
Pendrill, L.R. (2009).
\newblock {EURAMET: European Assoc. of National Metrology Inst.}
\newblock \emph{NCSLI Measure}, 4(4), 40--44.

\bibitem[{Poot et~al.(2022)Poot, Portegies, Mooren, van Haren, van Meer, and
  Oomen}]{Poot2022}
Poot, M., Portegies, J., Mooren, N., van Haren, M., van Meer, M., and Oomen, T.
  (2022).
\newblock {Gaussian Processes for Advanced Motion Control}.
\newblock \emph{IEEJ Journal of Industry Applications}, 21011492.

\bibitem[{Rasmussen and Williams(2006)}]{Rasmussen2004}
Rasmussen, C. and Williams, C. (2006).
\newblock \emph{{Gaussian processes for machine learning.}}
\newblock London, England.

\bibitem[{Takamasu et~al.(1996)Takamasu, Ozawa, and Asano}]{Takamasu1996}
Takamasu, K., Ozawa, S., and Asano, T. (1996).
\newblock {Basic concepts of nano-CMM}.
\newblock \emph{The Japan-China Bilateral Symposium on Advanced Manufacturing
  Engineering}, 155--158.

\bibitem[{Umetsu et~al.(2005)Umetsu, Furutnani, Osawa, Takatsuji, and
  Kurosawa}]{Umetsu2005}
Umetsu, K., Furutnani, R., Osawa, S., Takatsuji, T., and Kurosawa, T. (2005).
\newblock {Geometric calibration of a coordinate measuring machine using a
  laser tracking system}.
\newblock \emph{Meas. Science and Tech.}, 16(12), 2466--2472.

\bibitem[{Wegman(2006)}]{Wegman2006}
Wegman, E.J. (2006).
\newblock {Reproducing Kernel Hilbert Spaces}.
\newblock In \emph{Encyclopedia of Statistical Sciences}. John Wiley {\&} Sons,
  Inc., Hoboken, NJ, USA.

\bibitem[{Wu and Wang(2013)}]{Wu2013}
Wu, B. and Wang, B. (2013).
\newblock {Automatic measurement in large-scale space with the laser theodolite
  and vision guiding technology}.
\newblock \emph{Advances in Mech. Eng.}, 2013.

\end{thebibliography}

\end{document}